\documentclass{elsart}

\usepackage{amssymb, amsmath, graphicx, hyperref}

\newcommand{\mn}{\mathbb{N}} 

\newcommand{\Esym}{\mathrm{E}}
\newcommand{\E}[1]{\Esym\left[#1\right]}
\newcommand{\Probsym}{\mathbb{P}}
\newcommand{\Prob}[1]{\Probsym\left[#1\right]}

\newtheorem{theorem}{Theorem}

\begin{document}

\begin{frontmatter}

\title{Markov-modulated on/off processes for long-range dependent 
internet traffic}
 
\author{Richard G. Clegg}
\ead{richard@richardclegg.org}
\address{Department of Mathematics,
University of York,
York, UK.  YO10 5DD
}

\begin{abstract}
The aim of this paper is to use a very simple queuing model to
compare a number of models from the literature which
have been used to replicate the statistical nature of internet
traffic and, in particular, the long-range dependence of this
traffic.  The four models all have the form of 
discrete time Markov-modulated processes (two other models are
introduced for comparison purposes). 

While it is often stated that long-range dependence has a critical
effect on queuing performance, it appears that the models used
here do not well replicated the queuing performance of real
internet traffic.  In particular, they fail to replicate
the mean queue length (and hence the mean delay) and the 
probability of the queue length exceeding a given level.

\end{abstract}

\begin{keyword}
long-range dependence \sep Markov chains \sep Markov-modulated processes
\sep queueing \sep internet traffic
\end{keyword}
\end{frontmatter}

\section{Introduction}

Markov chains (MC) and Markov-modulated processes (MMP) are well-known 
modelling techniques which are successful in a wide variety of fields.  They
are also a traditional tool for queuing theory and for investigating networks
and queues of networks. 
In the last ten years several models have been introduced for 
the purposes of modelling internet traffic based on MMP.  These 
models are often motivated by the idea of capturing the
long-range dependence (LRD) which is seen in real internet
traffic and replicating the Hurst parameter $H$ which
characterises long-range dependence.  
The models have a common form, they
produce a process which is one or zero ({\em on} or {\em off}) and
work in discrete time.  The 
{\em on}/{\em off} process can then be seen as a time series of packets and
inter-packet gaps.  Obviously such a simplistic model of network
traffic cannot be expected to capture all the behaviour of a real
traffic network.  However, in exchange for this loss of realism,
the analytic simplicity of the models means that mathematical 
insights into their performance can be gained which might not
be seen in simpler models.  Additionally, the computational
performance of such simple models is typically good allowing
for fast modelling of large systems or large numbers of packets
in small systems.  

There is a considerable body of work on MMP to analyse the queuing of
internet traffic (\cite{li1993}, produces some useful results for
analysing queuing of MMP type models).  This paper
concentrates on those MMP models using {\em on/off} processes with the
intention of capturing LRD.
Four different MMP based models which were introduced with the
aim of modelling LRD 
are described in this paper.  Certainly many other models are possible.
In the case of one model (the pseudo-self-similar traffic model) some
new results (both theoretical and computational) are presented which
question whether this model can produce traffic with a known Hurst parameter.

The models are then compared in simulation with 100,000 packets from
two real traffic traces.  The traces are both easily available from
the internet for research.  The data sets chosen represent an older
but extremely well-studied packet trace (from 1989) 
and a more modern data set likely to be more representative of
modern internet traffic (from 2003).

It is found that not all the models examined can
reliably generate traffic with the required mean and Hurst parameter
(a standard measure of the degree of LRD present in data).  Even
when they do, the models fail to reproduce the queuing performance of the
real traffic in almost all cases.

\subsection{Long-Range dependence in internet traffic}

The introduction to LRD given here is, by necessity, brief.
For a fuller introduction see Beran \cite{beran1994} and
for an introduction in the context of internet traffic,
see Clegg \cite[Chapter one]{clegg2004}.  For a summary of work
on LRD in internet traffic see Willinger et al\cite{willinger2003}.

Let $X_t$ be a weakly-stationary time series $\{X_t:  t \in \mn\}$
with mean $\mu$ and variance $\sigma^2$.  The
autocorrelation function (ACF) as a function of lag $k$ is given by
\begin{equation*}
\rho(k) = 
\frac{\E{(X_t- \mu)(X_{t+k} - \mu)}}{\sigma^2}.
\end{equation*}

\begin{defn}
A weakly-stationary time series is said to be {\em long-range
dependent} (LRD) if the sum of its ACF is not convergent.  That is,
the sum $\sum_{k=0}^{\infty} \rho(k)$ diverges.  Note that sometimes
the weaker condition that $\sum_{k=0}^{\infty} |\rho(k)|$ diverges
is given.
\label{defn:lrd}
\end{defn}

Often a specific asymptotic form for the ACF is assumed,
\begin{equation}
\rho(k) \sim c_\rho k^{-\alpha},
\label{eqn:rho}
\end{equation}
where $k,c_\rho > 0$ are constants, $\alpha \in (0,1)$, and $\sim$ here
and throughout this paper means
asymptotically equal to as $k \rightarrow \infty$.  This form is
used to define the Hurst parameter which is given by $H= 1 - \alpha/2$.
Note that not all LRD processes necessarily have a definable Hurst
parameter but the Hurst parameter where $H \in (1/2,1)$ is usually
considered the standard measure of LRD.  For a discussion of measuring
the Hurst parameter in the context of LRD see \cite{clegg2006}.
Long-Range dependence is also sometimes expressed in terms of
asymptotic second order self-similarity.

A related topic is that of heavy-tailed distributions.
\begin{defn}
A random variable $X$ is heavy-tailed if, for all $\varepsilon > 0$
it satisfies
\begin{equation}
\Prob{X > x} e^{\varepsilon x} \rightarrow \infty \quad \mathrm{ as } \quad 
x \rightarrow \infty.
\label{eqn:exptails}
\end{equation}
\end{defn}
Again, often a specific form is assumed 
\begin{equation}
\Prob{X > x} \sim C x^{-\beta},
\label{eqn:tails}
\end{equation}
where $C > 0$ is a constant and $\beta > 1$.

From Heath et al \cite[Theorem 4.3]{heath1998}, heavy tails
and long-range dependence are related.  An
{\em on}/{\em off} process with heavy-tailed {\em on} periods of
the form given in \eqref{eqn:tails} and {\em off} periods which fall off
faster is a long-range dependent process.  Note that if
a process has heavy-tailed {\em off} periods and
{\em on} periods which fall off faster then this, then theorem
can still be applied since the ACF of an {\em on}/{\em off} process is
unchanged if {\em on} and {\em off} are reversed.

The area of long-range dependence became of interest to 
internet researchers when LRD was discovered in measurements
of packets per unit time on an Ethernet segment 
\cite{leland1993}.
It was later shown that the data sources for the traffic
exhibited heavy tails in their on periods \cite{willinger1997}.
These heavy tails are speculated to be the cause of the LRD in
internet traffic.
These measurements have been repeated many times since.  The
reason this is important is that LRD can have severe implications
for queuing performance.  Traffic exhibiting LRD can have much
longer delays although the relationship is not a simple one
\cite{neidhardt1998,sahinoglu1999}.

\subsection{Markov-modulated {\em on/off} processes}

A Markov-modulated {\em on/off} process can in general 
be written in terms of an underlying MC 
process
$\{X_t: t \in \mn\}$ (where $X_t$ is a discrete time homogeneous MC) 
and a 
derived process $\{Y_t: t \in \mn\}$ where 
\begin{equation*}
Y_t= 
\begin{cases}
1 & X_t \in A  \\
0 & \text{otherwise}, 
\end{cases}
\end{equation*}
where $A$ is some subset of the states of the chain.

\begin{figure}[htb] \begin{center}
\includegraphics[width=11cm]{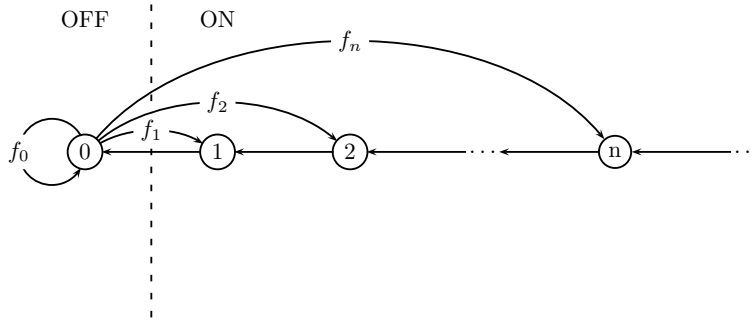}
\caption{A topology used by models which generate LRD 
for particular choices of $f_i$.}
\label{fig:cleggtop}
\end{center} \end{figure}

Because it is often desired that the model capture the LRD observed
in internet traffic, then the underlying MC for the process $X_t$
will usually be infinite.  The simplest example is probably the topology 
used by the Wang model \cite{wang1989} and the Clegg/Dodson model 
\cite{clegg2005} shown in Figure \ref{fig:cleggtop}.  The model produces
a series which is {\em on} 
(that is $Y_t = 1$ or alternatively the model emits a packet) 
when the underlying
MC (the process $X_t$) is in a non zero state.  Conversely it is {\em off}
($Y_t = 0$ or does not emit a packet) 
when the underlying MC is in the zero state.
A number of authors have used models of this form.  Four models
from the literature are discussed here.

Obviously this introduces a huge simplification into the
modelling.  The traffic model produced has uniform length packets
and which can only have arrival times 
$n\Delta t$ where $n \in \mn$ and $\Delta t$ is a characteristic
of the system.  Amongst those features not captured by such a model
are
\begin{itemize}
\item The distribution of packet lengths shown by real life systems.
\item The seasonality shown by real life systems (daily and weekly 
cycles).
\item System behaviour arising from TCP feedback mechanisms.
\item Usually packets are not constrained to arrive within some
multiple of $\Delta t$.
\end{itemize}
Set against this, there is the analytical simplicity of such models.
By investigating toy models where the behaviour can be understood
it may be possible to gain an insight which is not possible in 
more complex models which capture more realistic features of the
behaviour of network traffic.

\section{Traffic generation models considered}

Four models have been found in the literature of which two use
the same topology but different parameters.  In order to simplify
the explanations given in this paper, the models will be described 
using the same notation even where this will differ from the
notation given by the authors in the papers cited.  While the
Wang model was not actually suggested as a model for internet traffic,
it is included here as it is the oldest model the author has found
in the literature which associates on/off MMP with LRD.

The notation used is:
\begin{itemize}
\item ${\mathbf P}$ --- the transition matrix of the underlying MC.
\item $f_k$ --- the transition probability to some state $k$ (in the
models discussed it will usually be obvious where the transition was 
from).
\item $\pi_k$ --- the equilibrium distribution of some state $k$.
\item $\mu$ --- the mean number of packets generated per 
iteration ($\E{Y_t})$.
\item $\alpha$ --- a parameter giving the degree of LRD and related to
the Hurst parameter $H$ by $H= (2- \alpha)/2$.
\end{itemize}

The physical interpretation of these parameters is important to understand.
The interpretation of $f_k$ and $\pi_k$ depends on the topology of the
model.  For example, in the topology of the Wang or the Clegg/Dodson model 
then $f_k$ is the probability that a zero will be followed by 
exactly $k$ ones.  The probability $\pi_0$ is the probability
that (at equilibrium) the series will have the value zero --- hence
$\mu= 1-\pi_0$ is the mean number of arrivals per unit time for the model.  
Similar interpretations can be made for the Arrowsmith/Barenco model.

The value of $\mu$ is critically important since it controls the
amount of traffic the model will produce.  The parameter $\alpha$ (not
relevant for the PSST) which relates to the Hurst parameter also has important
effects for queuing.

\subsection{Wang model}

The Wang model \cite{wang1989} grew out of the problems associated
with calculating the invariant density of certain non-linear maps.
In particular, it arises from a piecewise linearisation of the
Manneville-Pomeau map \cite{pomeau1980} which is itself used
in internet traffic modelling \cite{erramilli1994}.  The topology of the MC
is given in Figure \ref{fig:cleggtop} and therefore the transition
matrix is
\begin{equation}
{\mathbf P}= \left[
\begin{array}{cccccc}
f_0 & f_1 & f_2 & \dots & f_n & \dots \\
1   &  0  &  0  & \dots &  0  & \dots \\
0   &  1  &  0  & \dots &  0  & \dots \\
0   &  0  &  1  & \dots &  0  & \dots \\
\vdots & \vdots & \vdots & \ddots & \vdots & \ddots
\end{array} \right]
\label{eqn:wangp}
\end{equation}
The finite version of this matrix is sometimes known as a 
{\em companion matrix}.
The MC can be shown to be ergodic if $f_0 > 0$, if the sum 
$\sum_{i=0}^\infty i f_i$ converges and if, for all $m \in \mn$ there is 
$n > m : f_n > 0$.  
For the Wang model, the transition probabilities are given by
\begin{equation*}
f_k=
\begin{cases}
\frac{a}{k^{\alpha+1}} - \frac{a}{(k+1)^{\alpha+1}} & k \in \mn  \\
1 - a & k = 0, 
\end{cases}
\end{equation*}
where $a \in (0,1)$ and $\alpha > 0$ are parameters of the model.  
It has been shown \cite{wang1989} that if $\alpha \in (0,1)$ then the
process will have LRD with 
\begin{equation*}
\rho(k) \sim k^{-\alpha}.
\end{equation*}
The mean $\mu$ is given for this topology (assuming the conditions for
ergodicity are met) by
\begin{equation}
\mu= 1 -\pi_0 = 1 - \left[1 + \sum_{k=1}^\infty k f_k\right]^{-1},
\label{eqn:mu}
\end{equation}
where the expression in brackets comes from the mean first return time
time for state 0.  For the other states it can be easily shown that
\begin{equation*}
\pi_k= \pi_0 \sum_{j=k}^\infty f_j.
\end{equation*}

Substituting into \eqref{eqn:mu} for the specific
values this becomes
\begin{equation}
\mu= 1 - \left[1 + \sum_{k=1}^\infty \frac{a}{k^{(\alpha +1)}} \right]^{-1}
= 1 - \left[1 + a \zeta(\alpha +1)\right]^{-1},
\label{eqn:wangmu}
\end{equation}
where $\zeta(\alpha+1)$ is the Riemann zeta function.
While this model allows the Hurst parameter to be set, there is no closed
form for the mean, though, given a value for $\alpha$ one could estimate
the correct value for $a$ by an iterative procedure.
It is easy to get expressions for the other equilibrium densities
\begin{equation*}
\pi_k= \frac{(1-a)\pi_0}{1 - \left[k/(k+1)\right]^{(\alpha+1)}} 
\qquad k \in \mn.
\end{equation*}

This model has the advantage that the LRD has been shown analytically and
can be set with a single parameter --- choice of $\alpha$ determines the
Hurst parameter of the simulated traffic.  Computational modelling is relatively
easy.  There is no closed form for the equilibrium density of any of
the states but the calculation is not difficult.  There is no
closed form solution for the value of $a$ corresponding to a particular mean
once $\alpha$ has been set, however, such a value can be approximated
using \eqref{eqn:wangmu} rearranged to
\begin{equation*}
a= \frac{\mu}{1 - \mu}
\zeta(\alpha+1)^{-1}.
\end{equation*}

\subsection{The Pseudo-Self-Similar Traffic (PSST) model}

\begin{figure}[htb] \begin{center}
\includegraphics[width=11cm]{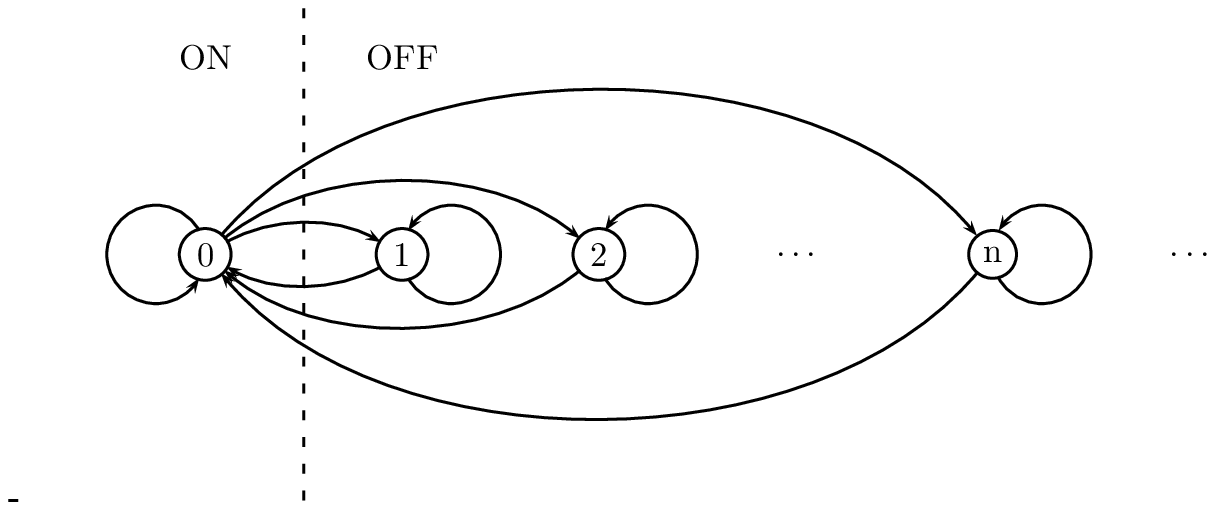}
\caption{The topology of the PSST model.}
\label{fig:psst}
\end{center} \end{figure}

The PSST model \cite{robert1997} was introduced to capture
the LRD in packet traffic (they use the phrase {\it self-similarity}).  
In fact the model suggested is a finite model which would not generate
self-similarity but the authors hope it would approximate it.  
The model is further investigated in \cite{khayari2004} and 
criticised as providing unrealistic
estimates for queuing performance.
The topology of the PSST model is shown in Figure \ref{fig:psst}.

The transition matrix for the model truncated after $n$ states 
(numbered $0$ to $n-1$) is given by.
\begin{equation*}
{\mathbf P}_n= \left[
\begin{array}{ccccc}
\Sigma_0 & \frac{1}{a} & \frac{1}{a^2} & \dots & \frac{1}{a^{n-1}} \\
\frac{q}{a} & \Sigma_1 & 0 & \dots & 0 \\
\left(\frac{q}{a}\right)^2 & 0 & \Sigma_2 & \dots & \\
\vdots & \vdots & \vdots & \ddots & \vdots \\
\left(\frac{q}{a}\right)^{n-1} & 0 & 0 & \dots & \Sigma_{n-1}
\end{array}
\right],
\end{equation*}
where 
\begin{equation*}
\Sigma_0 = 1 - (1/a) - (1/a^2) - \dots - (1/a^{n-1})
\end{equation*}
and
$\Sigma_i= 1 - (q/a)^i$ for $i = 1, \dots, n-1$.  Note that for
this to be a valid ergodic MC $a - a^{-(n-1)} > 2$ and $a > q > 1$.
In the original model
\begin{equation*}
Y_t=
\begin{cases}
1 & X_t = 0,  \\
0 & \text{ otherwise }, 
\end{cases}
\end{equation*}
This is a slightly strange choice because such a model would produce
runs of packets which have a length which decays exponentially
and runs of inter-packet gaps which are heavy-tailed.  In this
paper, this model will be referred to as PSST(a) and PSST(b) is
the same model with {\em on} and {\em off} reversed (that is $Y_t = 1$ if 
$X_t \neq 0$).

Previous references have used a truncated finite version of
this model.  However, there seems no particular reason to use
this approximation and here the infinite model will be used. 
For $\Sigma_0$ the sum becomes 
\begin{equation*}
\Sigma_0 = 1 - \frac{1 - (1/a)^{n-1}}{a - 1}
\end{equation*}
which for the infinite chain reduces to $\Sigma_0 = a/(a-1)$.

For PSST(a) model the mean is given by 
\begin{equation*}
\mu_a = \pi_0 = \frac{q^n - q^{n-1}}{q^n - 1},
\end{equation*}
which in the infinite model becomes
\begin{equation*}
\mu_a= \frac{q-1}{q}.
\end{equation*}
For the PSST(b) model the mean is given by $\mu_b = 1-\mu_a= 1/q$.
This can be rearranged to $q=1/(1-\mu_a)$ or $q= 1/\mu_b$.
The equilibrium probabilities of the states are given by
\begin{equation*}
\pi_k = \frac{\pi_0}{q^k}.
\end{equation*}

However, there is no obvious interpretation of the $a$ parameter which,
in some way, in combination with the ratio $q/a$,
controls the long term decay of the model.  Lower values of $q/a$ will lead to
longer sojourn times in higher numbered states as $q/a$ becomes smaller.  So it
might be expected that lower values of $q/a$ would lead to higher correlations
over large lags.  The long-term behaviour of the model is discussed in section
\ref{sec:ltpsst}.

\subsection{Arrowsmith/Barenco model}

\begin{figure}[htb] \begin{center}
\includegraphics[width=11cm]{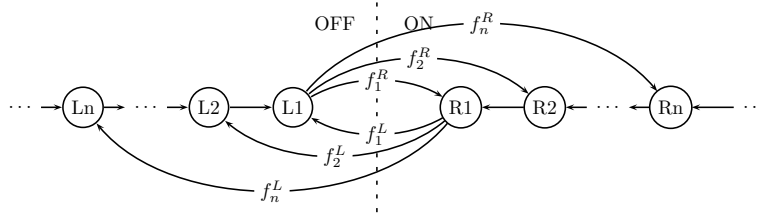}
\caption{The topology of the Arrowsmith/Barenco model.}
\label{fig:barentop}
\end{center} \end{figure}

This model \cite{barenco2002,barenco2004} was introduced to capture
the LRD seen in packet traffic and as a development of the
Wang model.  The topology is shown in Figure \ref{fig:barentop}.

The expected sojourn time on the left hand side of the model
$S_L$ is given by 
\begin{equation*}
S_L = \sum_{k=1}^\infty kf_k^L,
\end{equation*}
and a similar expression can be given for the sojourn time on the 
right hand side $S_R$.  This gives the expected number of packets
per iteration as
\begin{equation*}
\mu= \frac{S_R}{S_R + S_L}.
\end{equation*}

Because the model is a double-sided version of the Wang model
then the decay of consecutive runs of zeros and ones can be
individually controlled.  An important result with this
topology is \cite[Theorem 4]{barenco2004}.
\begin{theorem}
Let
\begin{align*}
f_k^L & \sim K_L k^{-\alpha_L+1} \\
f_k^R & \sim K_R k^{-\alpha_R+1}
\end{align*}
where $\alpha_L, \alpha_R > 0$ and $K_L, K_R > 0$ are all constants (here
we will restrict $\alpha_L, \alpha_R$ to $(0,1)$.
Then
\begin{equation*}
\rho(k) \sim K k^\beta,
\end{equation*}
where $\beta=\min(\alpha_L, \alpha_R)$ and $K$ is given by
\begin{equation*}
K=
\begin{cases}
\frac{K_R (1 - \mu)}{S(\alpha_R - 1) \mu} & \alpha_R < \alpha_L \\
\frac{K_L \mu}{S(\alpha_L - 1)(1 -\mu)} & \alpha_L < \alpha_R \\
\frac{K_R(1-\mu)K_L\mu}{\mu(1-\mu)S(\alpha_L -1)}
 & \alpha_L = \alpha_R,
\end{cases}
\end{equation*}
where $S= S_L + S_R$.
\label{thm:baren}
\end{theorem}

In fact, no specific form for the values of $f_k$ are given in 
\cite{barenco2004} and a variety of methods for choosing
parameters for this model are given in \cite{barenco2002}. 

\subsection{Clegg/Dodson model}

The Clegg/Dodson model \cite{clegg2005} uses the same topology as
the Wang model but different transition probabilities.  The model
has two parameters $\pi_0 \in (0,1)$ which is related to the mean
by $\pi_0 = 1 - \mu$ and $\alpha \in (0,1)$ which determines
the Hurst parameter.  These are
used to give the transition probabilities
\begin{equation}
f_k = \frac{1 - \pi_0} {\pi_0} \left[ k ^ {-\alpha} 
    - 2 (k+1) ^ {-\alpha}
    +(k+2) ^ {-\alpha}\right],
\end{equation}
for $k > 0$ and for $k=0$, 
\begin{equation}
f_0= 1 - \frac{1 - \pi_0} {\pi_0} \left[ 1 - 2^{-\alpha}\right].
\end{equation}

From these it can be shown that the model has equilibrium probabilities given
by 
\begin{equation}
\pi_k = (1 - \pi_0) [ k^ {-\alpha} - (k+1) ^ {-\alpha}] \qquad k > 0.
\end{equation}
which gives the sum
\begin{equation}
\sum_{i=k}^\infty \pi_i = (1 - \pi_0) k ^ {-\alpha} \qquad k > 0.
\label{eqn:piksum}
\end{equation}
This can be interpreted as the probability of a randomly chosen $Y_t$ being
one and being followed by at least $k-1$ ones.

The model is only valid with $\alpha, \pi_0 \in (0,1)$ if,
\begin{equation}
\pi_0 > \frac{2^\alpha - 1}{2^{\alpha+1} -1}.
\end{equation}
If this condition is not met the model does not form a valid Markov
chain.  This rules out combinations with high Hurst parameter and high
occupancy, near one (fortunately, these are unrealistic parameter
sets for most networks).

It can be proved that the time series $Y_t$ generated by this model 
exhibits LRD with the Hurst parameter $H= 1-\alpha/2$.

\subsection{Long-term behaviour of the PSST model}
\label{sec:ltpsst}

It has been speculated but not shown that the PSST model generates
traffic which exhibits long-range dependence.  Further, the traffic
from the PSST model has been analysed by measuring its Hurst parameter.  
However, there may be some reason to question this method.  
Consider $R_0$ the first return time to
the zero state of the model.  From the transition matrix it can be 
seen that,
\begin{equation}
\Prob{R_0 > k}  = \sum_{i=1}^n \frac{1}{a^i} (\Sigma_i)^k.
\label{eqn:R0simple}
\end{equation}
Since $\Sigma_i \in (0,1)$,
for any finite $n$ then as $k \rightarrow \infty$ this 
will fall off faster than some $x^k$ with $x \in (0,1)$
and therefore, this distribution cannot be heavy tailed.  Only the infinite
model can exhibit heavy tails in the return times to zero and hence the sojourn
time of zeros in the PSST or ones in the PSST(b) model.

It can be simply shown that the infinite model does have heavy tails in the
return time to zero.  Recall that
\begin{equation*}
\Sigma_i = 1 - (q/a)^i,
\end{equation*}
and since $q/a < 1$ then as $i \rightarrow \infty$ values of $\Sigma_i$ can
be found arbitrarily close to one.  To show that a distribution has a heavy tail
we must prove condition \eqref{eqn:exptails} holds for all $\varepsilon > 0$.
For a given $\varepsilon>0$ there must be some 
$N: \Sigma_N > e^{-\varepsilon/2}$ ($N$ is a function of $\varepsilon$ and
gets larger as $\varepsilon \rightarrow 0$).
Taking just the term for $i=N$ in \eqref{eqn:R0simple} gives
\begin{equation}
\Prob{R_0 > k} > \frac{1}{a^N} (e^{-\varepsilon/2})^k,
\end{equation}
and therefore, the condition for the distribution of $R_0$ being heavy tailed
is a condition on $\Prob{R_0 > k} e^{\varepsilon}$ and a lower bound is given
by
\begin{equation*}
\Prob{R_0 > k} e^{\varepsilon k} > \frac{1}{a^N} 
e^{-\varepsilon k/2} e^{\varepsilon k} = \frac{1}{a^N} e^{\varepsilon k/2}.
\end{equation*}
and for a fixed $\varepsilon > 0$ and $N$ (which is a function of $\varepsilon$) this
will become infinite as $k \rightarrow \infty$.
This shows that for the infinite PSST the length of
the OFF period has a heavy-tailed sojourn
time (or the ON period for the PSST(b) model).  However, the form of the asymptotic
fall-off may not be the often-assumed form of \eqref{eqn:tails}.  This would 
mean that theorem from \cite{heath1998} 
could not be applied and the LRD of the model could not be assumed.

The expression can be further expanded to give
\begin{align*}
\Prob{R_0 > k} & = \sum_{i=1}^n \sum_{j=0}^k \binom{k}{j} 
(-1)^j \left( \frac{q^j}{a^{j+1}} \right)^i \\
&= \sum_{j=0}^k  \binom{k}{j} 
\frac{(-q)^j(1 - q^{jn}/a^{(j+1)n}) }{(a^{j+1} - q^j)}.
\end{align*}
As $n \rightarrow \infty$ then $q^{jn}/a^{(j+1)n} \rightarrow 0$ and
this becomes
\begin{equation}
\Prob{R_0 > k} = \sum_{j=0}^k  \binom{k}{j}
\frac{(-1)^j }{a(a/q)^j - 1}.
\label{eqn:rosum}
\end{equation}
For $k$ odd the series can also be written as
\begin{align*}
\Prob{R_0 > k} & =  \\
 \sum_{j=0}^{(k-1)/2} \binom{k}{2j} & \left[\frac{1}{a(a/q)^{2j} - 1}
- \frac{1}{a(a/q)^{(k-2j)} -1} \right].
\end{align*}

While these expression are in a closed form they are not
particularly convenient to work with computationally.  
The binomial coefficient becomes
large for large $k$ as does the value of $(a/q)^j$ for large $j$ this
makes the two expressions above hard to work with numerically and an arbitrary
precision arithmetic library must be used to investigate large values of $k$.

\begin{figure}[htb] \begin{center}
\includegraphics[width=11cm]{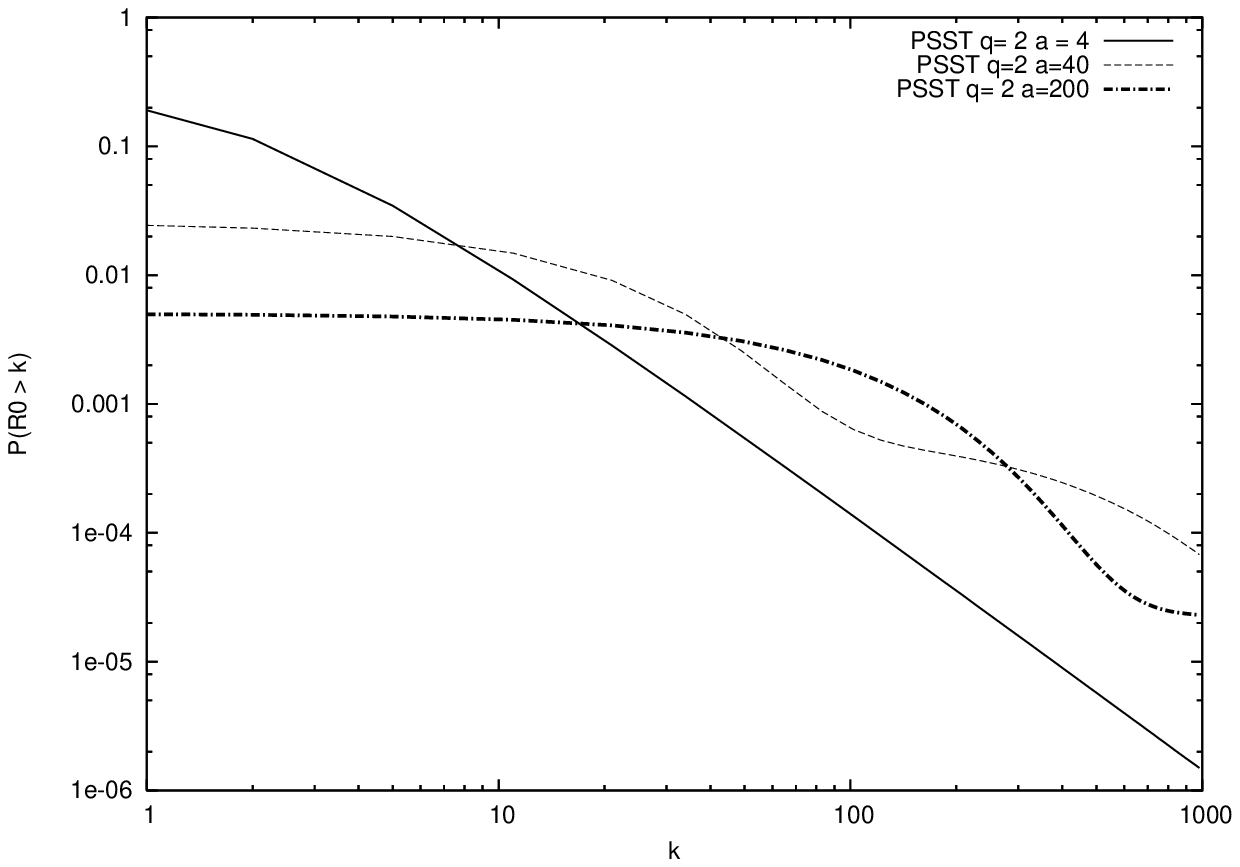}
\caption{Plot of $\Prob{R_0 > k}$ versus $k$ for PSST model.}
\label{fig:psst_pdf}
\end{center} \end{figure}

Figure \ref{fig:psst_pdf} shows three different plots for \eqref{eqn:rosum}
which all produce the same mean traffic (controlled by the $q$ parameter) but
vary the $q/a$ ratio which controls the dependence.  If the system has a 
power-law sojourn time then the log-log plot would tend to
a straight-line as $k \rightarrow
\infty$.  However, as can be seen, this is not the case for two of the
three lines here and for the range of $k$ plotted 
(although arguably might be the case for the $q/a=1/2$ line).
These plots look at sojourn times of up to 1000.  It may be that the 
plot would become linear.  However, the case where $k = 1000$ is the case 
where 1000 or more consecutive zeros (or ones in the PSST(b) model) 
appear and must be at the limits of what would be expected in even exceptionally
long computational runs.  In other words, this is the region of the model which
computational runs will ``see".

In short, while the PSST model has a heavy-tail distribution in the sojourn time
of the {\em off} 
state, it does not appear to have the expected power-law fall off.  The
author knows of no theorem which would prove the LRD or otherwise of 
the generated
time-series.  Even if the time-series generated did exhibit LRD it could be expected
that it is the case that the LRD would appear in the summation of 
the autocorrelation of Definition \ref{defn:lrd}
but not necessarily with the power-law fall of of \eqref{eqn:rho}.  If this were
the case then the model would not have a well-defined Hurst parameter.  
Measurements of the Hurst parameter on traffic produced by the PSST model 
vary greatly depending on the estimator used.  This is discussed in 
section \ref{sec:Hurst} and does not seem to have been noted by other
investigators.

\subsection{Computer implementation}

These MMP described are relatively simple and fast to implement.  For this paper,
they have been implemented in the computer language python.  By necessity
a robust computational implementation of such routines must deal with 
very small probabilities.  The phenomenon of LRD generated by such chains
relies on the robust computation of the probability of some long run of
ones or zeros.  However, numerical rounding issues become important 
in practical implementations and a naive implementation would be subject
to rounding errors.  The rounding error problem is described here in 
relation to the Clegg/Dodson model.  The full details are given in 
\cite{clegg2005}.  A robust implementation of any of these models is
must overcome similar rounding problems and a similar approach can be used.

For example, a direct computation of $f_k$ for 
large $k$ in models using the topology \ref{fig:cleggtop} would lead to
a very small number which may be hard to deal with computationally.
Assume that
the chain is in state zero at time $t$ ($X_t = 0$).  It is
now necessary to find $\Prob{X_{t+1}=k}$ using the values of $f_k$.
A naive implementation would choose the state to move to from the
zero state by picking a single random number and comparing it with
the probability distribution function.  For low $k$ a direct computation
of $\Prob{X_{t+1}=k}$ is well within the computational accuracy
of the computer.  For higher values then calculations of 
$\Prob{X_{t+1} \in [k,2k] | X_{t+1} \geq k}$ keep the numbers
in a manageable range.  This technique is employed for all four
models here.

It should be noted that LRD is a difficult subject to work with
computationally.  Hurst parameters near unity ($\alpha$ near zero) cause
great problems.  Theoretically, the nearer one the Hurst parameter,
the slower the sample mean will converge to the mean.  To give a 
concrete example, $10^6$ iterations of the Clegg/Dodson model
with $\mu=0.5$ and $\alpha=0.1$ ($H=0.95$) gives for the
first three samples, $\overline{Y}= 0.563$, $\overline{Y}= 0.401$ and 
$\overline{Y}= 0.426$.  This does not indicate a problem with the
computational model rather this slow convergence of the sample mean
is inherent in the nature of LRD itself.  Naturally, this has a
critical importance to real experiments.  

\section{Experimental setup}

The experiments performed in this paper are extremely simple.  The input data for
one simulation is a set of arrival times and packet lengths.  These data may
originate from real measurements or from the models described. 
The packets are then simulated as arriving at a queue of known output bandwidth.
The properties of the queue and the output traffic are then measured.  The
experiment may then be repeated with a smaller output bandwidth to see how this
affects the queue.  Obviously as bandwidth decreases it would be expected
that the mean queue length and queuing delays would increase but the exact
behaviour depends upon the statistical nature of the traffic.

\subsection{Data sets used}

Two data sets are used for the simulation in this paper.  In both cases, 
only the first 100,000 packets were investigated.  The names and origin
of the exact sources used are given here so that other researchers can make
similar measurements.

CAIDA data: This data set is taken from a trace approximately an
hour long.  It is referred to as {\tt 20030424-000000-0-anon.pcap.gz}
and was captured on
the 24th April 2003.  It
was captured on an OC48 link with a rate of 2.45 Gb/s.  The average
packet length was 493 bits.  The data is freely available to researchers
who fill in a request form.  More
information about this data can be found at: \\
\url{www.caida.org/data/passive/}.

Bellcore data: This much-studied data set is described in \cite{leland1991}. 
The data here is taken from an August 1989 measurement referred to as
{\tt BC-pAug89.TL}.
The data was collected
on an Ethernet link which connected a 
LAN to the outside world.  Note that in
this case the data did not record the true length of packets, only the
length less the Ethernet header (which is variable).  The average
packet length recorded was 464 bits. 
Hence, the experiment in this paper is only using an approximation 
of the real data.  The data is freely available for researchers.
More information about this data can be found at: \\
\url{ita.ee.lbl.gov/html/contrib/BC.html}.

\subsection{Queuing model, pre-processing and digitisation of real data}

The queuing model used in this paper is extremely simple.  The system
has a given bandwidth $b$ (bits/sec).  Items join the back of the queue.
If a packet of length $L$ bits arrives at the queue then it will take a time
$L/b$ to process.  It is output from the queue at this time.  Until this time
the entire packet is considered to be part of the queue for purposes
of calculating mean queue length (which can be calculated in terms of 
packets or bits).

A starting point for modelling is to establish a base case for comparisons.
The real data simply consists of arrival times of packets and packet lengths.
In order to attempt to match this data with real models, then a bandwidth $b$
was selected.  This was chosen to create an occupancy near ten percent as this
was thought to be a reasonable occupancy for a congested network.  The Bellcore
data was reported as being taken from a network with an occupancy of twenty
percent at peak times.  The CAIDA data almost certainly had an occupancy much
lower than this since it is from a modern high-speed link.  The actual figure chosen
is not really important since the data are then to be queued through lower and lower
bandwidths.  

For the Bellcore data the baseline bandwidth was chosen as
1.96Mb/sec and for the CAIDA data 128Mb/sec this gave occupancies of and 0.094 
and 0.098 respectively.  Traffic with the recorded arrival times and packet lengths 
was then passed through this queue and the output times from this queue were taken
as the base case to simulate.  The data referred to as ``raw" for the rest of this
paper is the output of this queue with either the Bellcore or CAIDA packet 
lengths as an input.

The traffic generation models are all {\em digitised} in a way that the real data was
not.  The models all simply produce a string of ones and zeros corresponding to a packet
or a gap.  To convert these
into packets and departure times a timescale $dt$ must be established and also a fixed
packet length.  The timescale $dt$ is the length of time between packets in a packet train
or the length of one inter-packet gap.  The packet length $l$ bits
was chosen as the mean packet 
length of the real data (464 bits for the Bellcore data and 496 bits for the CAIDA).  
The timescale was then
chosen related to the bandwidth as the time taken to transmit one
packet of this length, that is $dt = l/b$. 

Obviously this is a considerable simplification and it is therefore useful to 
investigate to what extent the real data would be altered if it were subject to
this {\em digitisation}.  Therefore a {\em digitised} 
version of the real data was 
produced where all the packets are of length $l$ and broadcast at 
fixed multiples of $dt$.  This
data is referred to as the digitised data.  It was created simply by simulating 
a queue where the packets arrival times and packet lengths were taken from the
real data.  At every time $n dt: n \in \mn$, if the queue contained $l$ or more bits
then a packet of length $l$ was sent from the queue at this time.  The data
referred to as {\em digitised} 
throughout the rest of this paper is the results
of this process with the input as the raw data and queued using
the same bandwidth $b$ as the raw data.

\begin{figure}[htb] \begin{center}
\includegraphics[width=11cm]{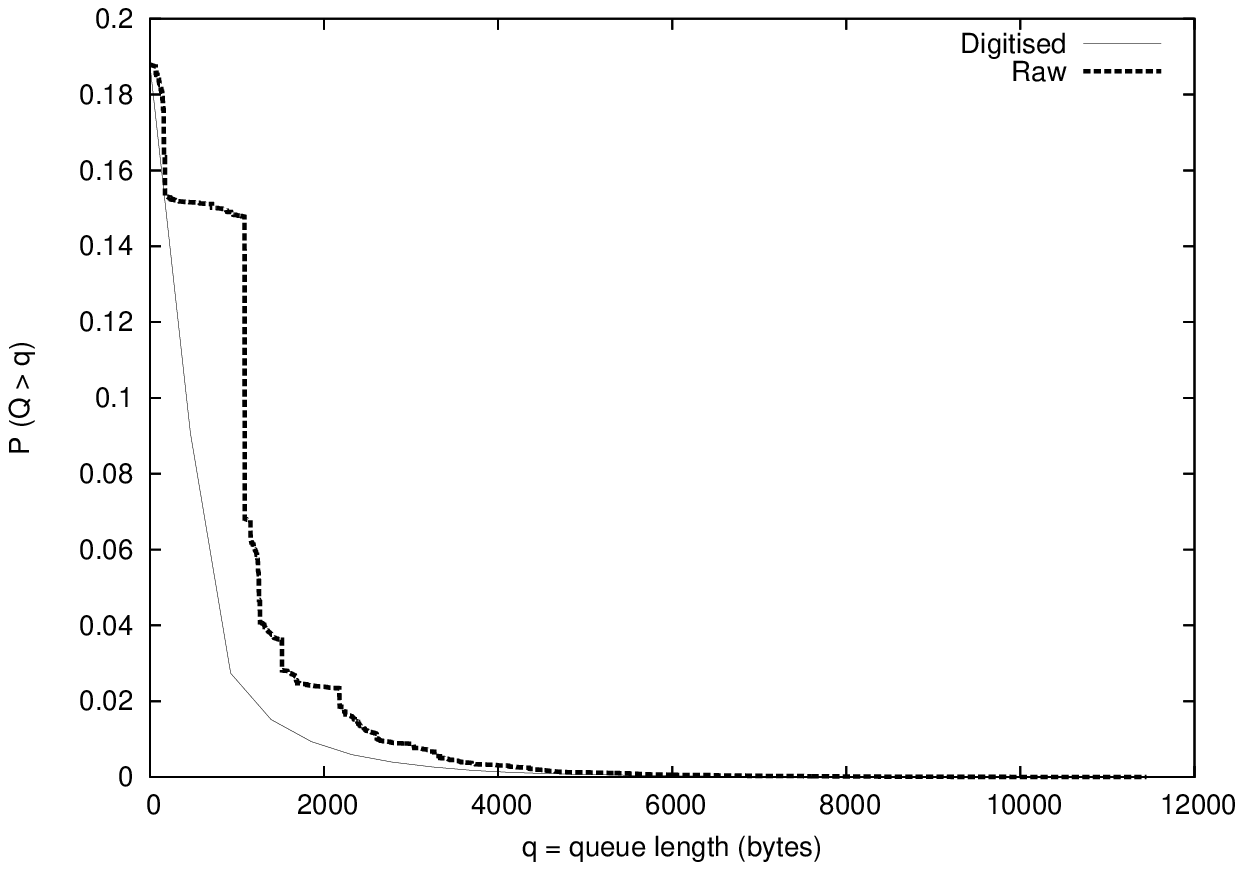}
\includegraphics[width=11cm]{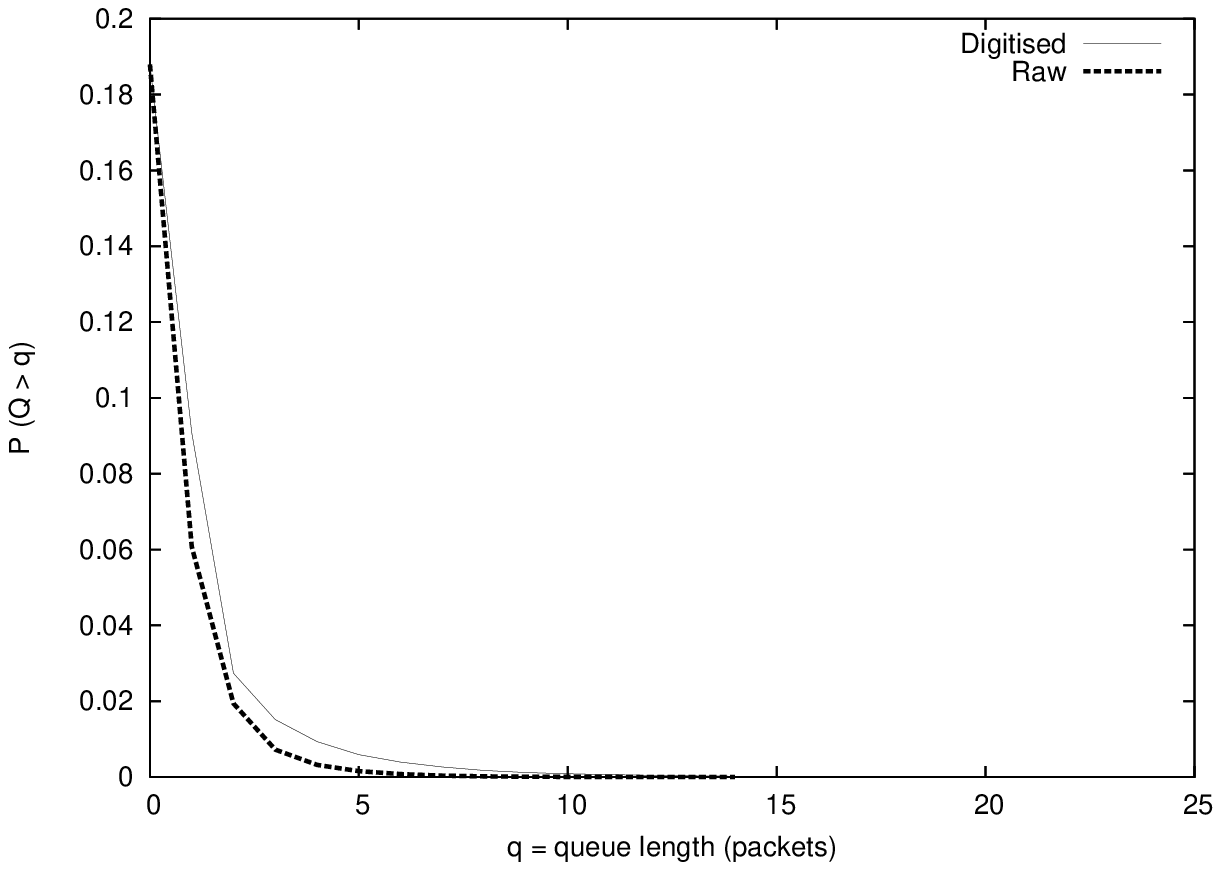}
\caption{Plot of $\Prob{Q > q}$ with $q$ as the queue length in bits(top)
and packets(bottom).}
\label{fig:dig}
\end{center} \end{figure}

Figure \ref{fig:dig} shows the differences introduced
by this digitisation process.  These results are produced by queueing the Bellcore
raw and digitised data in a queue with half the original bandwidth ($b$ is reduced
from 1.96Mb/s to 0.98Mb/s).  The top figure shows the  
distribution of the queue size in bits and the bottom figure shows
the distribution of the queue in packets.  As can be seen on the data in bits, the
real data has a much more complex graph, simply because packets can have a variety
of different lengths.  Interestingly the raw data tends to have a higher queue
length in terms of bits but lower in terms of packets.  The reason for this
is not known.  The digitised data certainly shows differences to the raw data
but the queuing performance is not greatly dissimilar.  Further comparisons
will be shown in the next section.

\subsection{Models used to simulate data}

Several {\em on/off} 
type models were tested against the real data to compare queuing
performance.
In addition to the MMP models already described two further models were used 
for comparison.  The Poisson model is, perhaps, the simplest possible model.  It
either generates or does not generate a packet in every time period with a
probability equal to the required mean.  Note that strictly speaking this is
a Bernoulli process but it is well approximated by a Poisson process.
Fractional Gaussian Noise (FGN) is a well known process for
generating data series with LRD.  While it produces continuous traces it can
be simply adapted to produce an {\em on/off} packet-gap model 
In practice, the PSST model was found to be extremely unstable for producing
traffic with the low mean used.  Successive realisations produced traces with
extremely different means and the results from the model varied greatly from
run to run --- for example, with $q=1.104$ the model should produce a mean of
0.94.  With $a=20$ three subsequent runs produced sample means of 0.11, 0.19 
and 0.27.  With $a=2$ (the lowest valid value for $a$)
three subsequent runs produced sample means of
0.11, 0.11 and 0.07.  In other words, the traffic level can increase by a huge
amount between runs and it is impossible to get consistent results.
For this reason, only the PSST(b) model results are presented here as this
model produces a more consistent sample mean.

For the Arrowsmith/Barenco model a method was needed for estimating the $f_i^L$ and
$f_i^R$ parameters.  These parameters represent the probabilities of an inter-packet
gap or packet train of length $i$ and this gives an obvious strategy for
tuning them.  The parameters are matched to the probability distribution of
inter-packet gap and packet train lengths in the digitised data for each sample
of traffic.  This implies that the Arrowsmith/Barenco model will be producing traffic
traces with the same distribution of packet train lengths and inter-packet gaps
as the digitised data.  Another method is described in \cite{barenco2004} where the
parameters of the model are tuned using genetic algorithms
to match a given autocorrelation function.  Many possible parameter selection 
strategies are available for this model and the comments in the rest of this paper
should only be taken as reflective of the particular method of parameter selection.
Note that this method for getting parameter settings means that the model does
not replicate the Hurst parameter.

The Poisson model is a single parameter model, reflecting only the mean of the data.
The Wang, Clegg/Dodson and FGN models are two parameter models which model the
mean and the Hurst parameter.  The PSST(b) model is a two parameter model which
models the mean and a parameter related to correlations.  The Arrowsmith/Barenco
model as used here is a multi-parameter model which models the probability 
distribution of the lengths of packet trains and inter-packet gaps.

\subsection{Calculation of Hurst parameter}
\label{sec:Hurst}

Calculating the Hurst parameter of real data is not a simple matter.  For 
a practical guide in the context of telecommunications and descriptions of the
methods used in this paper see \cite{clegg2006}.  The methods used
in this paper are the R/S estimator, Aggregated Variance, Periodogram, 
Wavelet analysis and the Local Whittle Estimator.
Software using the statistics package R can be downloaded from: \\
\url{http://www.richardclegg.org/lrdsources/software/}.  
An excellent description of
various methods and S-Plus code can be found at: \\ 
\url{http://math.bu.edu/INDIVIDUAL/murad/home.html}.

To calculate the Hurst parameter the data must first be converted into a
time series.  This can be done simply by counting the number of bytes processed
if the data is split into sample times of a given period.  
The question is then which time period to choose.  Too
long a time period will give too small a sample size to work with (a 
ballpark figure
is that several thousand time series points is a minimum).  Too short
a time period will give a time series which largely consists of zeros.

For the Bellcore date time periods of 0.1 seconds, 0.01 seconds and 0.001 seconds
were tried.  This resulted in 2521, 25208 and 252080 samples respectively.  However, 
at the smallest sample size more than three quarters of the periods sampled have
no packets at all.

\begin{table}[ht]
\begin{center}
\begin{tabular}{|l| l l l l l|} \hline
Data & R/S & Agg. V. & Period. & Wav. & Loc. W. \\ \hline
Raw (0.1s) &
0.757  &  0.509 & 0.657  &  0.689  &  0.855  \\
Raw (0.01s) &
0.75  &  0.751  &  0.845  &  0.769 &  0.828  \\
Raw (0.001s) &
0.798  &  0.765  &  0.79  &  0.809   &  0.736  \\ \hline
Dig (0.1s) &
0.756  &  0.509  &  0.657  &  0.689   &  0.856  \\
Dig (0.01s)  &
0.748  &  0.751  &  0.845  &  0.769  &  0.828  \\
Dig (0.001s) &
0.787  &  0.765  &  0.788  &  0.81  &  0.79  \\ \hline
\end{tabular}
\end{center}
\caption{Hurst parameter estimates for Bellcore data.}
\label{tab:hurstbell}
\end{table}

Table \ref{tab:hurstbell} shows the results for the raw and digitised
Bellcore data for the five estimators and three sampling rates as described
above.  The value of $H$ would normally be expected to lie between 0.5 (independent
or short-range dependent data) and 1.  As can be seen, at the slowest sampling
rate (smallest sample size) there is little agreement between the estimators.  
It would be hard to justify giving more than one significant
figure for $H$ given the low agreement between the estimators.  When 
sampling at the two higher rates the estimators seem to agree on a Hurst parameter
of around 0.8 (the exception being the Local Whittle estimator which gives
0.736 at the highest sampling rate for the raw data). 
The digitisation has made very little difference to the estimated Hurst parameter 
of the data in almost all cases.

The same procedure is performed for the CAIDA data.  This data has a line speed
and aggregation levels of 1000, 100 and 10 micro seconds were chosen which gives
3924, 39331 and 392301 samples respectively on the raw data (as with the Bellcore
data, the raw and digitised data were in broad agreement and the digitised results 
are not presented here.)  The results are shown in Table \ref{tab:hurstcaida} and,
again, while it is only appropriate to get $H$ to one decimal place the results broadly 
agree with a value of $H=0.6$.  This is a low level of LRD (indeed it may be there
is no LRD in this data) consistent with the rule of thumb that networks with
lower utilisation often exhibit a lower level of long-range dependence.
\begin{table}[ht]
\begin{center}
\begin{tabular}{|l| l l l l l|} \hline
Data & R/S & Agg. V. & Period. & Wav. & Loc. W. \\ \hline
Raw (1000us) & 0.567  &  0.617  &  0.563  &  0.641  &  0.621  \\
Raw (100us) & 0.66  &  0.593  &  0.607  &  0.639   &  0.635  \\
Raw (10us) & 0.529  &  0.626  &  0.644  &  0.578   &  0.529  \\ \hline
\end{tabular}
\end{center}
\caption{Hurst parameter estimates for CAIDA data.}
\label{tab:hurstcaida}
\end{table}
\begin{table}[ht]
\begin{center}
\begin{tabular}{|l| l l l l l|} \hline
Data & R/S & Agg. V. & Period. & Wav. & Loc. W. \\ \hline
FGN (0.1s) & 0.76  &  0.713  &  0.85  &  0.874  &  0.84  \\
FGN (0.01s)  & 0.754  &  0.779  &  0.798  &  0.763  &  0.793 \\
FGN (0.001s) &  0.652  &  0.781  &  0.786  &  0.724   &  0.605  \\ \hline
Wang (0.1s) & 0.634  &  0.426  &  0.381  &  0.502  &  0.68 \\
Wang (0.01s) &0.603  &  0.563  &  0.653  &  0.675  &  0.853  \\
Wang (0.001s) & 0.608  &  0.747  &  0.802  &  0.851  &  0.898  \\ \hline 
PSST (0.1s) & 0.553  &  0.481  &  0.451  &  0.589  &  0.629  \\
PSST (0.01s) & 0.796  &  0.582  &  0.617  &  0.693  &  0.972  \\
PSST (0.001s) & 0.705  &  0.709  &  0.868  &  1.07  &  1.35 \\ \hline
Arr./Bar. (0.1s) & 0.553  &  0.504  &  0.458  &  0.448  &  0.549  \\
Arr./Bar. (0.01s) & 0.634  &  0.52  &  0.538  &  0.597  &  0.619  \\
Arr./Bar. (0.001s) & 0.582  &  0.59  &  0.605  &  0.612  &  0.703  \\ \hline
\end{tabular}
\end{center}
\caption{Hurst parameter estimates for 100,000 packet realisations of
simulated data.}
\label{tab:hurstsim}
\end{table}
Hurst parameter estimates were also made for the simulated data.  The Wang model
and FGN models were uses to simulate the Bellcore data using a theoretical
Hurst parameter $H=0.8$ 
and the same theoretical mean as the Bellcore data.  The models are 
run to produce 100,000 packets and aggregated as before to produce Hurst parameter
estimates.  The PSST(b) model is also used with the same theoretical mean
as the data and a=500.

Table \ref{tab:hurstsim} shows the Hurst parameter estimates for
models attempting to simulate the parameters  of the Bellcore data.  As can be seen,
despite the theoretically sound nature of the FGN and Wang method, the
estimators are not actually in very good agreement with the theory.  This is
not an unfamiliar situation to researchers studying the field of LRD estimation.
From the table, the FGN Hurst parameter is estimated reasonably at an
aggregation level of 0.1 and 0.01 seconds and rather underestimated at an aggregation
level of 0.001 seconds.  The Wang model also produces a variety of estimates
for $H$ with the R/S estimator being particularly bad and the estimates at
an aggregation of 0.1 seconds being so low that a researcher might conclude
there was no LRD present.

The PSST model is spectacularly inconsistent.  Considering the $H$ parameter is
usually expected to be in the range $H=(1/2,1)$ for LRD and $H=1/2$ for no LRD, the 
estimates for the model vary across the entire available range and outside it.  The
estimates also change completely depending on the aggregation level considered.  This
is consistent with the hypothesis that if the PSST model does indeed produce
traffic with LRD it does so in such a way that the traffic has no Hurst parameter.

Note that the method for choosing parameters for the Arrowsmith/Barenco model
does not reproduce the Hurst parameter and this can be seen.  While the Bellcore
data had a measured Hurst parameter around $H=0.8$ the Arrowsmith/Barenco
model tuned to have the same distribution of packet train lengths and inter-packet
gaps has a measured Hurst parameter around $H=0.6$.

If the sample trace is longer then better estimates of the Hurst parameter are
obtained for all models apart from the PSST(b) model.  Table \ref{tab:hurstsim2} 
shows a typical set of results for 1,000,000 packet realisations.  The Wang model
is producing data with a Hurst parameter of 0.8 and, while all of the estimators
but the R/S are overestimating, they are not doing so greatly and are largely
consistent with scales.  The Clegg/Dodson model and FGN model also behave as
expected.  The PSST(b) model continues to produce widely differing estimates
for $H$ which vary with aggregation scale.  The PSST model shows similar behaviour.

Note that \cite{khayari2004} reports that the PSST model was tuned to replicate
the value of $H$ for real data.  It is possible that the authors only had a single 
$H$ estimator available and thus did not notice these discrepancies.  In this
paper the $a$ parameter for the PSST(b) model was chosen simply to be ``large"
for the high Hurst case and smaller for the low Hurst case.  No more scientific
fitting procedure was available for the model.
\begin{table}[ht]
\begin{center}
\begin{tabular}{|l| l l l l l|} \hline
Data & R/S & Agg. V. & Period. & Wav. & Loc. W. \\ \hline
Wang (0.01s) & 0.612  &  0.885  &  0.858  &  0.842  &  0.875 \\
Wang (0.001s)  & 0.724 & 0.859 & 0.840 & 0.852 & 0.905\\ \hline 
PSST(b) (0.01s)  & 0.604  &  0.668  &  0.682  &  0.761  &  0.971 \\
PSST(b) (0.001s) & 0.846  &  0.74  &  0.88  &  1.076  &  1.349  \\ \hline
\end{tabular}
\end{center}
\caption{Hurst parameter estimates for 1,000,000 packet realisations of simulated 
data.}
\label{tab:hurstsim2}
\end{table}
\section{Results}

\label{sec:results}
The experiments performed are all of the same nature.  The input to an experiment
is data either from a real data source (raw or digitised) or from one of the
models with its parameters tuned to match that of the digitised data.  The input
data is then sent through a queue with a given bandwidth $b$.  The queuing
performance of the model is then measured.  While many performance measures
could be considered, results are only given here in terms of the expected queue
size $\E{q}$.  The experiment is then repeated with a smaller value of $b$ until 
experiments have been performed with occupancies ranging from 0.1 to 0.6 (the 
latter representing a network with an extremely high degree of congestion).

\subsection{Bellcore data}

All models were run to produce traces 252 seconds long (the length of the original
trace) with packets of length 464 bits.  
The model parameters were all chosen to replicated the mean of the
original data.  For the Clegg/Dodson, FGN and Wang models the second model parameter
was chosen to replicate the Hurst parameter $H=0.8$ (although parameters higher and
lower were tried).  For the PSST(b) model $q = 10.4$ was chosen to capture the correct
mean.  Various values of $a$ were tried but, as has been mentioned, tuning the
model to replicate the Hurst parameter was impossible.  The results presented here
used $a=500$ which was considered to provide quite a high level of correlations.  
The model was tested with various other $a$ values with little more success than that
reported here.  The Arrowsmith/Barenco model was tuned to replicate the distribution
of the packet-train lengths and inter-packet gaps of the digitised data as previously
described.

\begin{figure}[htb] \begin{center}
\includegraphics[width=11cm]{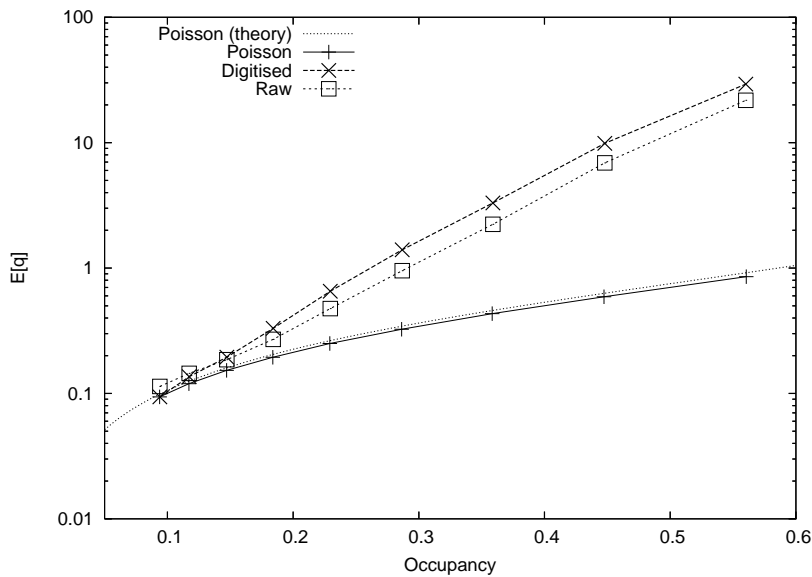}
\caption{Comparison of Poisson model versus 
real traffic for Bellcore data.}
\label{fig:poisvsreal}
\end{center}
\end{figure}

\begin{figure}[htb] \begin{center}
\includegraphics[width=11cm]{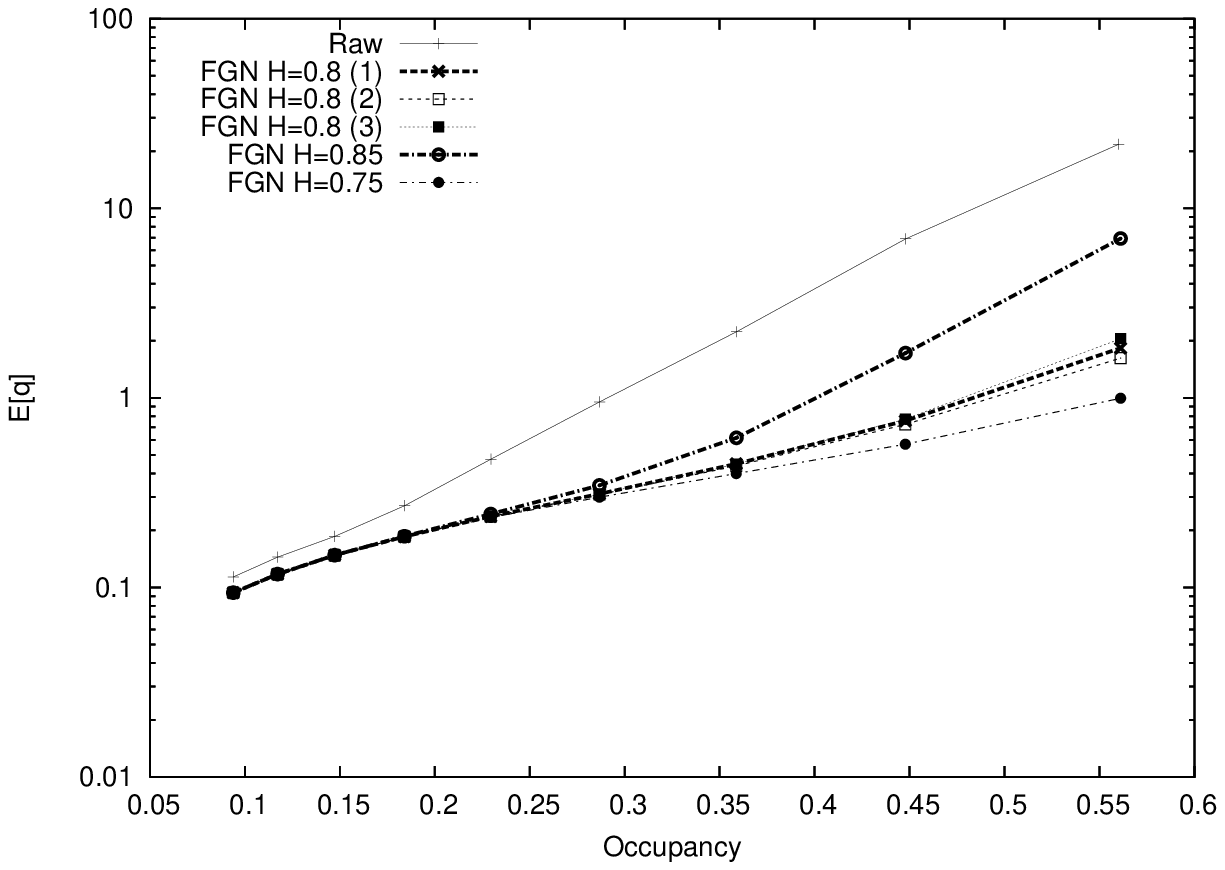}
\caption{Comparison of FGN model versus real traffic for Bellcore data.}
\label{fig:fgnvsreal}
\end{center} \end{figure}

\begin{figure}[htb] \begin{center}
\includegraphics[width=11cm]{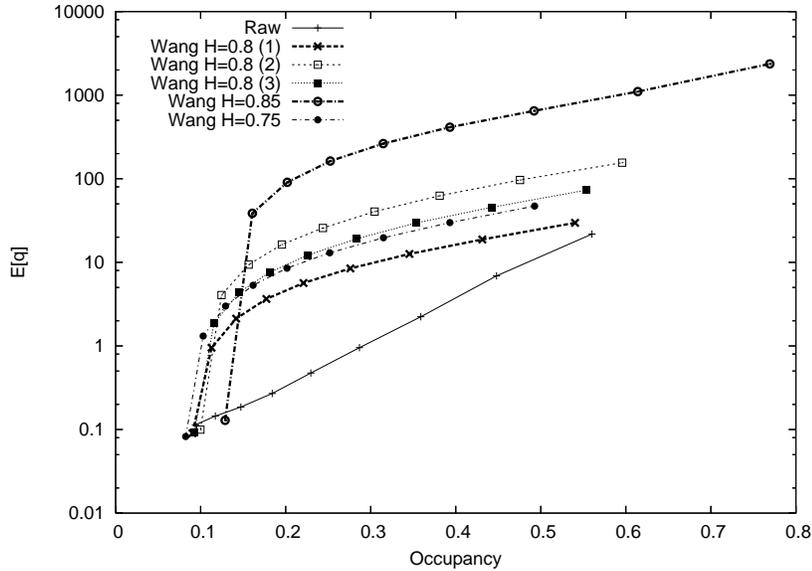}
\caption{Comparison of Wang (model versus 
real traffic for Bellcore data.}
\label{fig:wangvsreal}
\end{center} \end{figure}

Figure \ref{fig:poisvsreal} shows comparisons 
of the Poisson model with the real data (both raw and digitised).  Note that
the $y$ axis is a logscale on this and all following figures. The theory line for the
Poisson model is provided by the Pollaczek--Khinchin formula (with the
discrepancy being accounted for by the fact that the model is, strictly
speaking, Bernoulli not Poisson).  As can be seen, and as would be expected, 
the Poisson model hugely underestimates the level of queuing in the network.
This result is as would be expected from the literature.

Figure \ref{fig:fgnvsreal} 
shows a comparison of the FGN model versus real traffic.
Several realisations of FGN are tried.  Three realisations have $H=0.8$, one has
$H=0.85$ and one has $H=0.75$.  As can be seen the model produces more or less
the same queuing performance with the same mean and Hurst parameter and a higher
Hurst parameter produces larger queues.  This is as would be expected from the
literature.  However, both are under predicting the queuing of the real data.

Figure \ref{fig:wangvsreal} 
shows a comparison of the Wang model versus real traffic.
Again several realisations of the Wang model are used.  Again the model is producing
similar queuing performance for the three models with the same mean and Hurst parameter.
Note that the model with a Hurst parameter of $H= 0.85$ has higher occupancy simply
because this realisation happens to have produced data with a sample mean greater
than the actual mean of the process.  This is a common problem with LRD processes with
high $H$ where the sample mean can converge slowly to the actual mean.  The most 
striking thing though is that again all the models have failed to capture the 
queuing performance of the real data.  In this case the models have greatly over
estimated the amount of queuing which will occur.

\begin{figure}[htb] \begin{center}
\includegraphics[width=11cm]{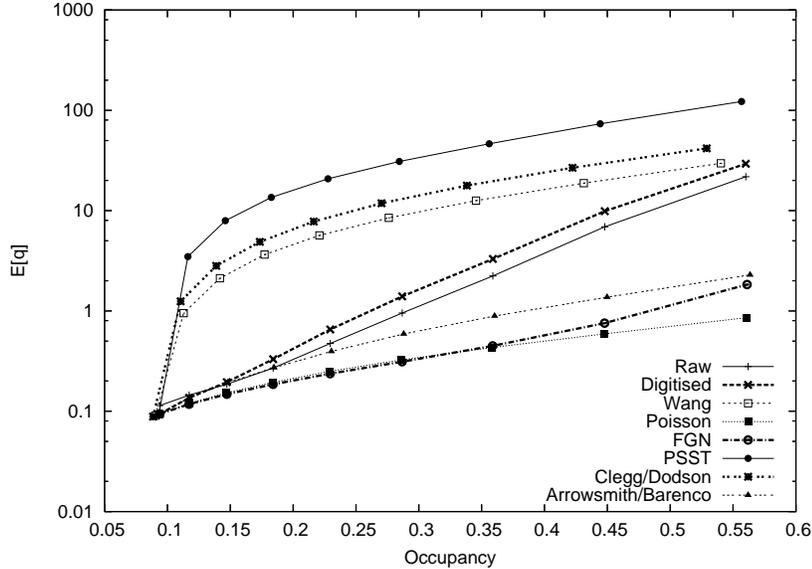}
\caption{Comparison of all models on Bellcore data looking at the expected
queue length.}
\label{fig:bellcoreall}
\end{center} \end{figure}

Figure \ref{fig:bellcoreall} shows a comparison with all of the models used
against the real traffic trace.  What is most striking about this
is that none of the models are even close to replicating the real data.  The
raw and digitised data are relatively close together.  The Clegg/Dodson
and Wang models appear to be similar in performance (perhaps unsurprisingly
since they have the same topology but different parameters).  Both of these models
overestimate queuing.  The PSST model
produces a higher queue level than these two models and is, obviously an overestimate
of the queuing of the real data.  The Poisson model, as has been mentioned, is an
underestimate of the real queuing performance as is the FGN model.  Interestingly,
for low occupancies, the Poisson model is actually giving higher queues than
the FGN model even though this model was motivated by addressing the underestimation
of queuing in the Poisson model.  The Arrowsmith/Barenco model is, perhaps, the closest
model to the real data but this particular method must still be regarded as having
failed to successfully model the queuing performance of the Bellcore data.  Also the
model used is a multi-parameter model as opposed to a one or two parameter model
like the others and hence would be expected to be a much closer match.

A subtle but important difference in the figure is that in the regions with
higher occupancy (the right hand side of the graph) the slope of the lines
is very different.  With the exception of the FGN model, in this region the models
appear to have parallel lines on this figure but these lines have a very different
gradient to the plots for the raw and digitised data.  In other words, not only
the level of congestion is different but the way the data responds to an increase
in congestion is fundamentally different.

\begin{figure}[htb] \begin{center}
\includegraphics[width=11cm]{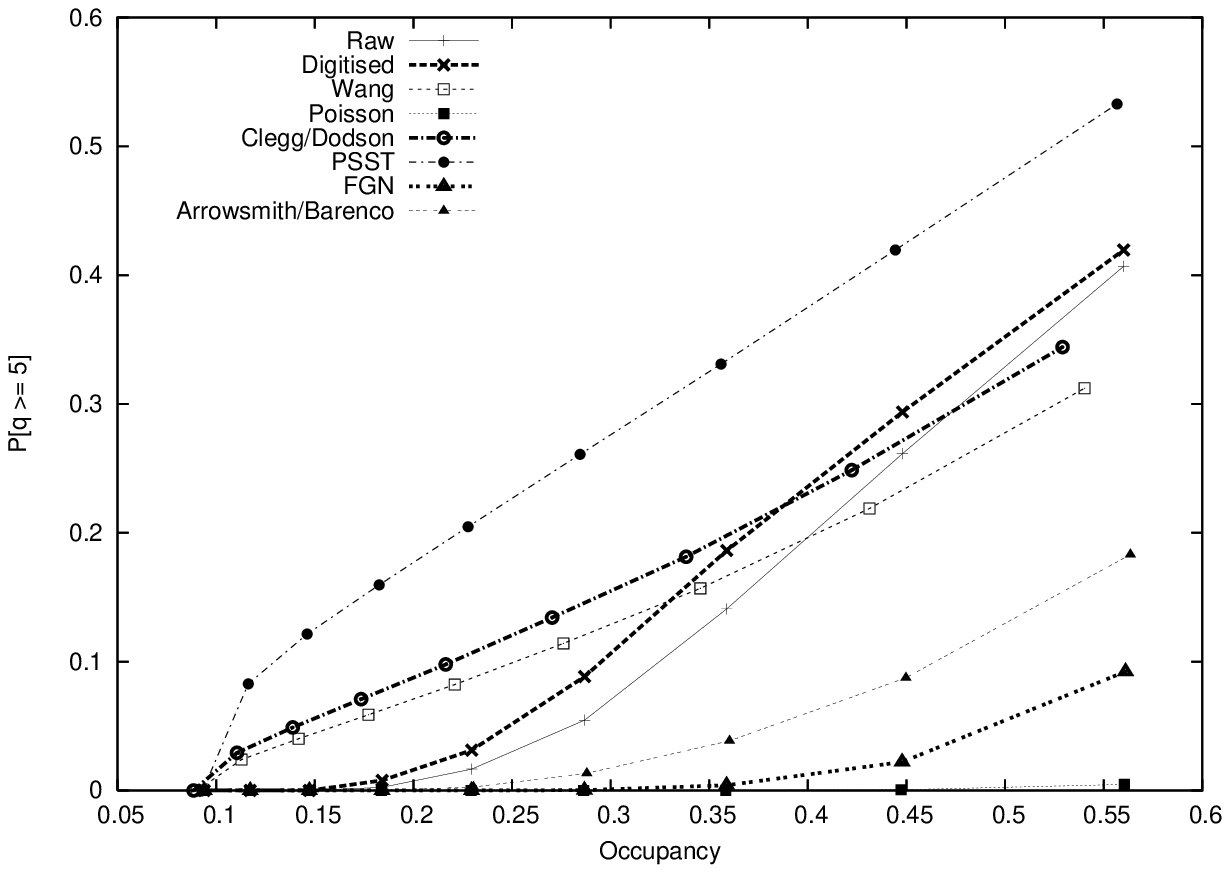}
\includegraphics[width=11cm]{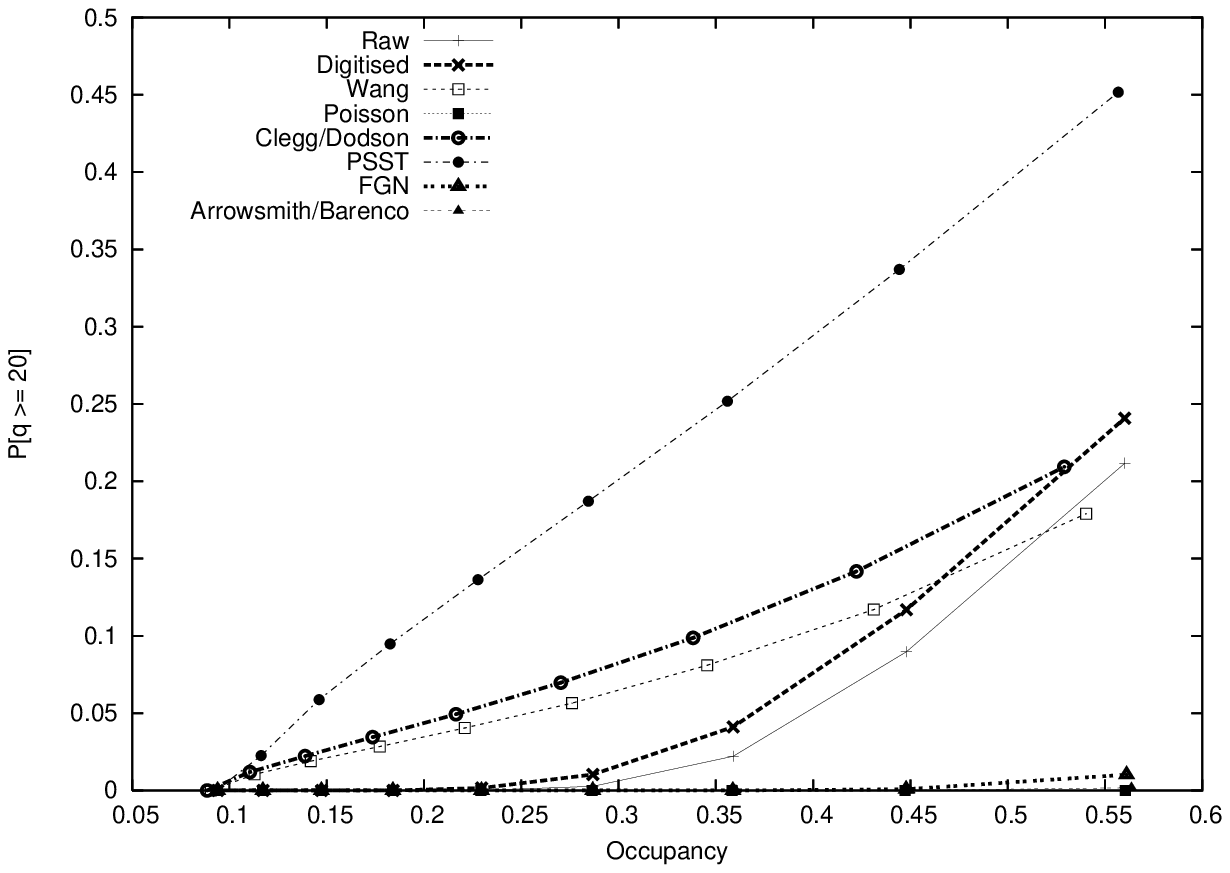}
\caption{Comparison of all models on Bellcore data looking at the 
probability that the queue is five or greater (top) or twenty or
greater (bottom).}
\label{fig:bellcoreover}
\end{center} \end{figure}

Another way of comparing the models is to look at the probability of given
queue lengths.  Here, a similar graph to Figure
\ref{fig:bellcoreall} is plotted with the $y$ axis as the probability
of the queue being equal to or greater than a given length.  Figure
\ref{fig:bellcoreover} shows the probability of the queue equalling or
exceeding five (top) or twenty (bottom).  As can be seen, again
none of the models are doing a good job of approximating these probabilities.
The raw and digitised data remain similar to each other.  At low occupancies
the Poisson, FGN and Arrowsmith/Barenco models seem to be the best approximations
and at high occupancies the Clegg/Dodson and Wang models seem to be closer.
None of the models prove accurate over the whole range.  The models prove
poor approximations over the whole range of occupancies considered,
no matter what the queue length chosen.

\subsection{CAIDA data}

All models were run to produce traces 4.02 seconds long (the length of the original
trace) with packets of length 496 bits.  
Again the model parameters were all chosen to replicated the mean of the
original data and, for the Clegg/Dodson, FGN and Wang models the Hurst parameter 
$H=0.6$.  For the PSST(b) model $q = 10.2$ was chosen to capture the correct
mean and $a=30$ was used to give a low level of correlation.  The Arrowsmith/Barenco
model was tuned as for the Bellcore data.

\begin{figure}[htb] \begin{center}
\includegraphics[width=11cm]{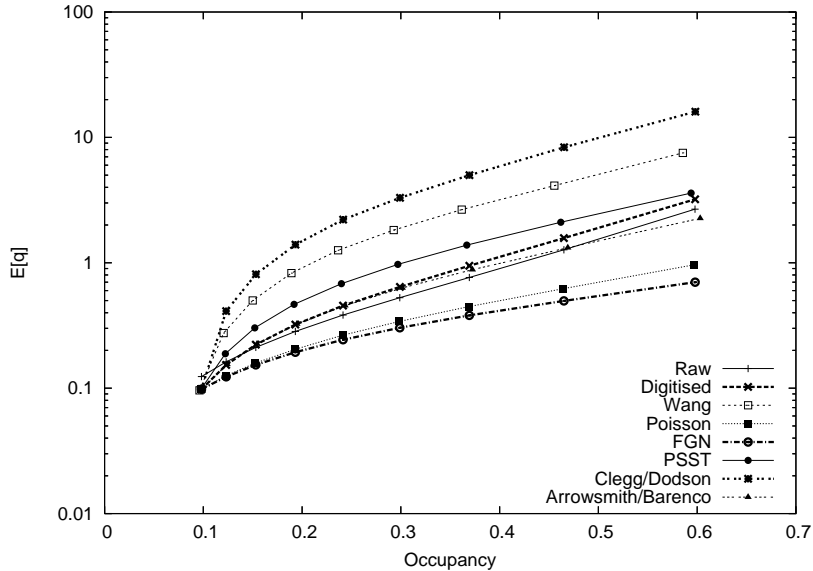}
\caption{Comparison of all models on CAIDA data.}
\label{fig:caidaall}
\end{center} \end{figure}

Figure \ref{fig:caidaall} shows a comparison of all the models versus the real and
digitised data.  In most ways the results are similar to the results of modelling
the Bellcore data.  Again the Clegg/Dodson and Wang models are similar but provide
an overestimate of the level of queuing.  Again the FGN and Poisson models provide
an underestimate of the queuing in this case with the FGN model giving a lower
estimate of queuing than the Poisson model.  
In this case, however, the Arrowsmith/Barenco model has provided a very good 
estimate of the queuing lying somewhere between the raw and digitised data.
The PSST model has provided a better approximation although it is still an over
estimate.  Again, however, the same feature can be seen as with the Bellcore data,
in the high occupancy region (at the right hand side of the graph) the artificial 
models (with the exception of the FGN) seem to have run parallel (they appear
to have approximately the same gradient).  However, the real data appears to
have a steeper gradient than any of the models in this region.

\begin{figure}[htb] \begin{center}
\includegraphics[width=11cm]{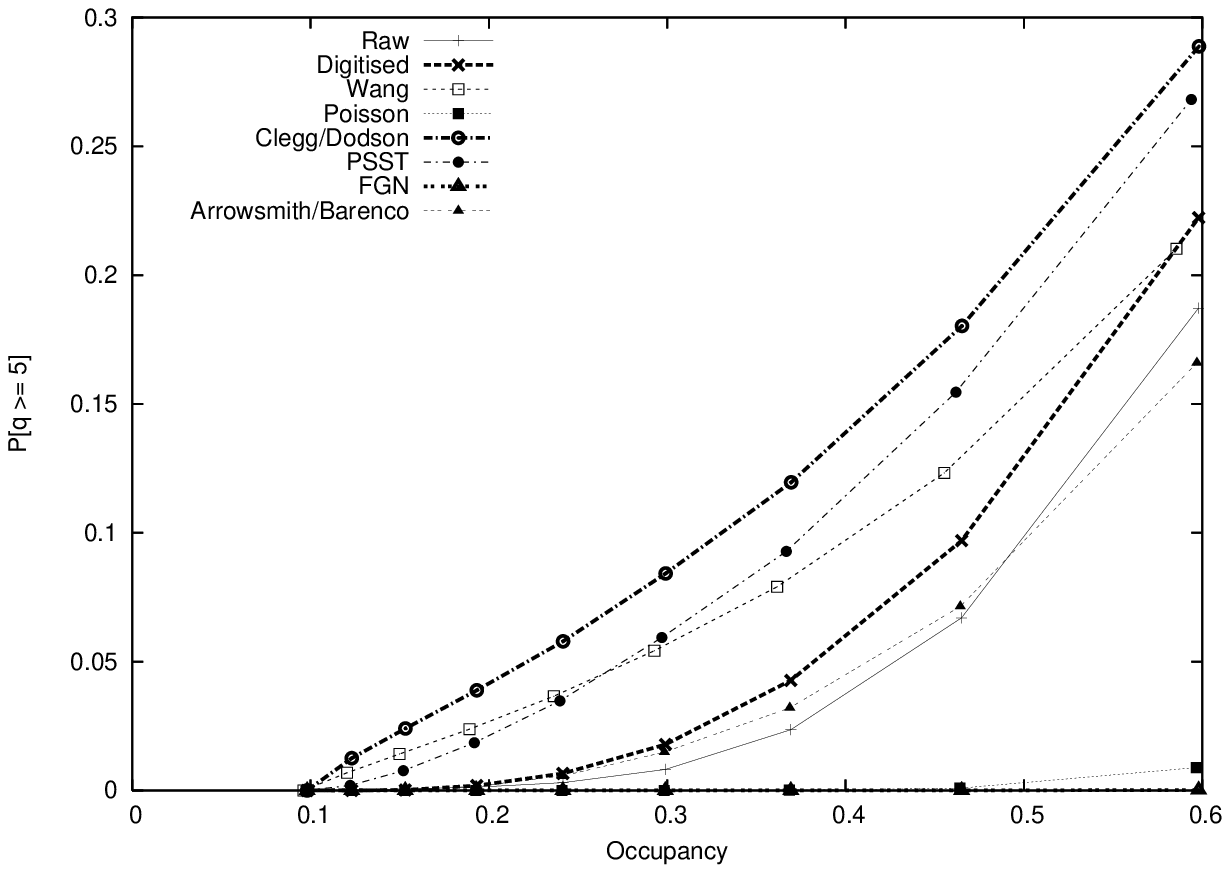}
\includegraphics[width=11cm]{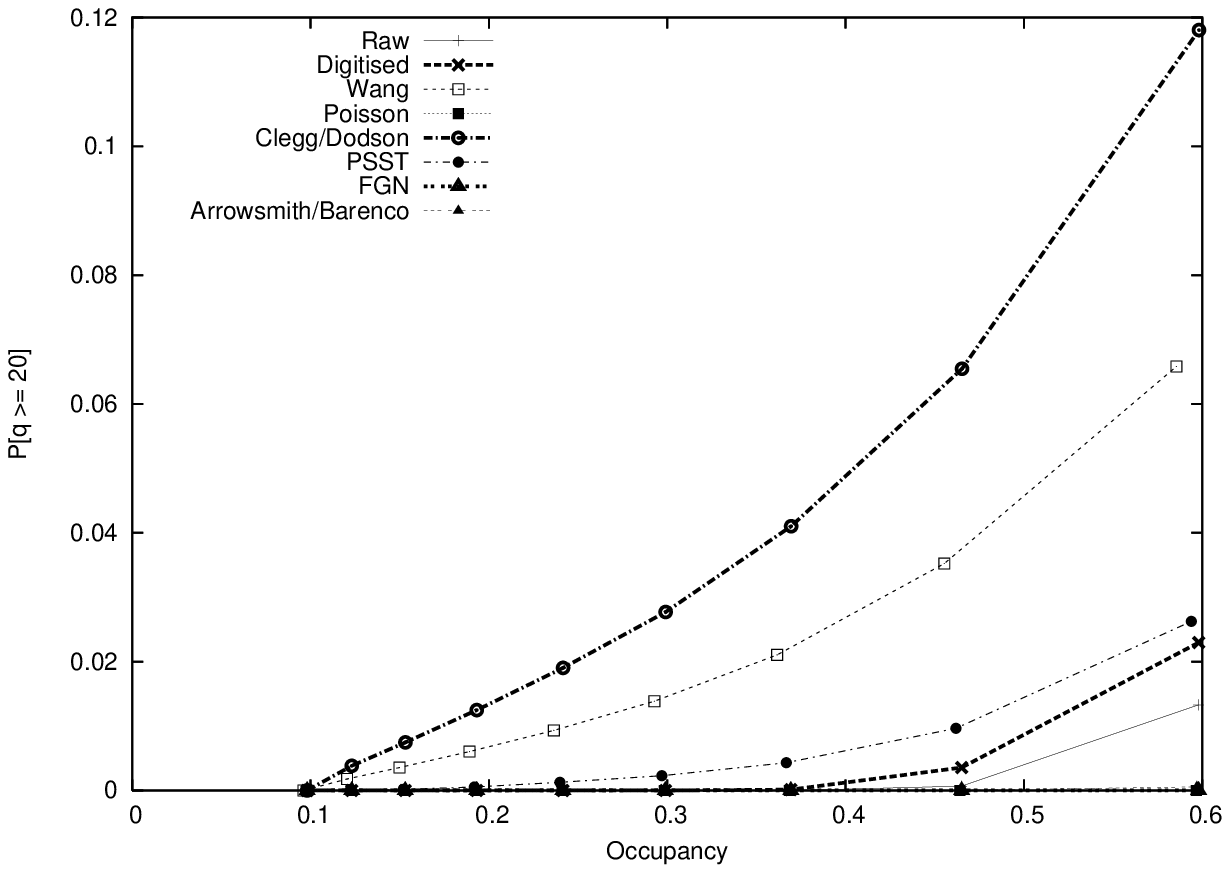}
\caption{Comparison of all models on CAIDA data looking at the 
probability that the queue is five or greater (top) or twenty or
greater (bottom).}
\label{fig:caidaover}
\end{center} \end{figure}

Again, the experiments can be repeated looking at the probabilities of
the queue exceeding a given length.  As for the Bellcore data, the 
probability of the queue length exceeding certain levels are plotted.
Figure \ref{fig:caidaover} shows the probability that the queue equals
or exceeds five (top) or twenty (bottom).  In the case of the queue
exceeding five, the Arrowsmith/Barenco model gives an excellent approximation
for most of the range of the experiment.  Indeed it lies between the raw
and digitised lines.  The Wang and Clegg/Dodson (and possibly even PSST) models
could also be seen to be acceptably close.  For the more extreme event
assessed by the probability that the queue exceeds twenty, the Wang and 
Clegg/Dodson models are clearly overstating the probability of these 
extreme queues.  The PSST(b) model is a relatively good approximation of
the real data.  All other models seem to be underpredicting the 
likelihood of large queues at high occupancies.  

\subsection{Later sections of data}

\begin{figure}[htb] \begin{center}
\includegraphics[width=11cm]{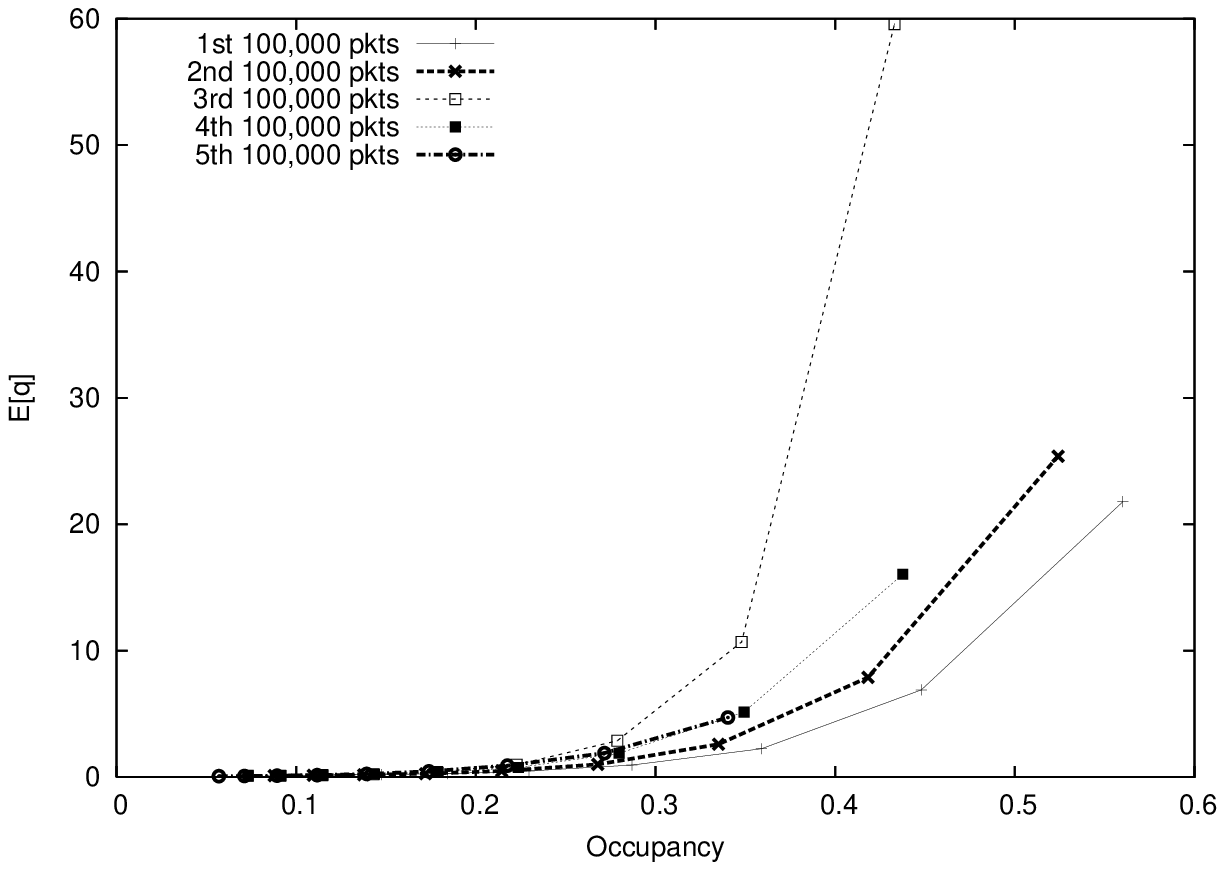}
\includegraphics[width=11cm]{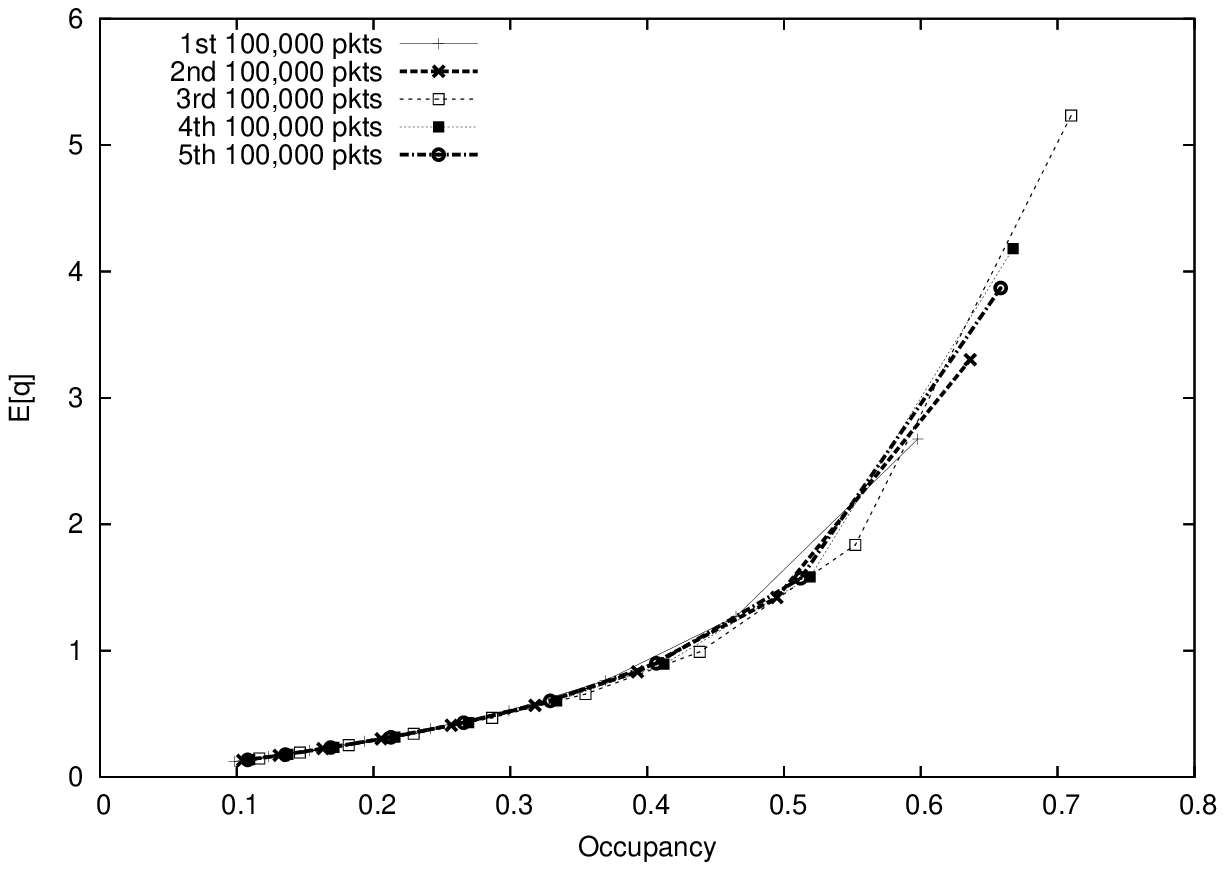}
\caption{Comparison of subsequent sections of Bellcore data (top) and CAIDA data
(bottom).}
\label{fig:next}
\end{center} \end{figure}

As has been seen, it is a difficult task to replicate the queuing performance of
a sample of 100,000 packets of real data.  It might then be asked, if this sample
could, in principle be replicated with one hundred percent accuracy then would this
modelling be appropriate for subsequent data samples from this data.  Figure
\ref{fig:next}(top)  shows the second, third, fourth and fifth samples of 100,000 packets
of the Bellcore data
queued by the same process (raw data).  As can be seen the queuing performance of
the data varies greatly between these samples.  Note that each point plotted
corresponds to a different bandwidth for queuing but the data differs also in the
mean packets transmitted and hence the occupancy differs between samples.  
However, the differences are far from being differences purely due to the
time over which the packets transmitted.  Consider, for example the first 100,000 packets
compared to the third 100,000.  The third 100,000 packets have a lower occupancy 
(that means they took longer to transmit and are sampled from a period of time where,
on average, packets were being sent at a lower rate) but a higher queue. 

Figure \ref{fig:next} (bottom) shows a similar plot for the CAIDA data.  In this 
case, while the mean rate of data transmission still differs between samples,
the queuing performance is broadly similar.

\subsection{Discussion and criticism of results}

The results presented here show an important weakness in a class of MMP
models which have been used to emulate network traffic.  Over all the data
sets used, no models gave a good representation of the expected queue or
queue overflow probabilities for the Bellcore data.  
It should be noted again that the model referred to
as the Arrowsmith/Barenco model was just one method for choosing parameters for
this model and this is not a general criticism of using this topology for 
modelling data.  For the expected queue length, 
the Arrowsmith/Barenco model gave a good representation
of the CAIDA data and the PSST(b) model gave a fair representation. In the case
of the PSST(b) model there is no good way the author knows of picking one of
the model's parameters and it may be that this match is little more than 
coincidence.  No models seemed able to predict the queue overflow probabilities
of the CAIDA data for all queue lengths and all occupancies (although perhaps
this would be demanding too much of such simple models).

The CAIDA data had a lower degree of long-range dependence.  
Also the CAIDA model
had a more consistent performance between subsequent samples of 100,000 packets.
It may be suspected that these two facts are related but it is hard to 
tell without further investigation.

One obvious criticism of the experiments performed here is that a real
network would not behave like this under queuing.  The TCP/IP protocol
incorporates mechanisms which perform crude congestion control.  In short,
what is described here as the real data is not, in fact, how a real network
would perform subject to the capacity constraints.  This criticism is an 
important one and it is certainly true that a closed loop model incorporating
this feedback would be a better representation of what actually happens when
a real network becomes more congested.  However, this said, good open loop
models would greatly help the understanding of what factors in real 
network traffic impact on queuing performance.  It would be much harder to
understand how bandwidth impacted a closed loop system and it could be 
reasonably expected that changes to a network which were positive for
an open loop system were also positive for a closed loop system (although
this is by no means guaranteed).

Other models may be capable of producing a better model of the queuing performance
of internet traffic.  In \cite{barenco2002} a method is described for tuning
the parameters of the Arrowsmith/Barenco model to replicate the ACF of a traffic
sample using genetic algorithms.  Wavelets have been used not just for analysis
of traffic traces (as in this paper) but also for simulating traffic \cite{riedi1999} 
\cite{abry2003}.  It is not clear, however, how an individual packet model could be
generated from a time series produced using wavelet.  

\section{Conclusion}

This paper presented a number of MMP models which produce {\em on/off}
sequences.  These models have all been suggested as potential models of
telecommunications data (with the exception of the Wang model which arose in
the field of Statistical Mechanics).  In addition the FGN model was included 
since this is also a commonly suggested model for telecommunications data and
the Poisson model was included as a baseline for comparisons.  It is clear
that in all cases the Poisson modelling was inappropriate and gave misleading
queuing estimates.

While the Wang, Clegg/Dodson and FGN models captured the mean and Hurst 
parameter of the data, they did not accurately reflect the queuing 
performance of the system.  Investigation of those models made it clear 
that, within those models, the Hurst parameter had a very important 
effect on queuing performance with a high Hurst
parameter equating to worse queuing performance.

The PSST model proved hard to work with and the criticisms of this model by
other authors seem justified \cite{khayari2004}.  
However, this author is sceptical of the claim in \cite{khayari2004} that
the model can be fitted to provide a particular Hurst parameter.  The PSST model 
does not seem to provide a traffic trace with a controllable Hurst parameter.  In
addition the original PSST model exhibits a remarkable degree of instability in 
its sample mean when the mean is set to produce a low level of traffic (a value of
$q$ near 1).  This said, the PSST(b) model was the only two parameter model to 
have any degree of success in modelling either data set.

The Arrowsmith/Barenco model as used in this paper did not model the Bellcore data
well but was a very good model for the CAIDA data.  It should be recalled that in
this paper, a particular method for fitting the model parameters was used and the
Arrowsmith/Barenco model is more general than the particular model used here.  
The model used in this paper is a multi-parameter model and would be expected 
to provide a better fit than one or two parameter models.

It is important to note that in the case of the high Hurst parameter Bellcore 
data, even if a model were found which accurately represented the queuing performance
of the first 100,000 packets subsequent samples of 100,000 packets behaved very 
differently.  On the other hand, for the CAIDA data which had a lower Hurst parameter,
subsequent samples of the data performed more consistently.  

In short, the problem of replicating the statistics of real traffic traces is a
difficult one.  The models tried in this paper all have their attractions in 
terms of computational or mathematical simplicity but none of them proved adequate
to model the queuing performance of a traffic trace taken from a real network.
If researchers are to truly understand what causes and can relieve queuing in 
telecommunications networks then an open loop model of this type would be an
important starting point.

\bibliography{rgc_cn2006}


\end{document}